# Solar cell Edge shunt isolation : A simplified approach

I. M. Hamammu[*] and K. Ibrahim
School of Physics
University Sains Malaysia
11800 Pinang, **MALAYSIA**
Email: *Hamammu@yahoo.com*
*Permanent address Faculty of Science
Garyounis University,
P.O.Box 9480, Benghazi, **LIBYA**

*Abstract*: **Edge shunts are among the main factors causing reduction in solar cell open circuit voltage that is why their isolation is one of the main steps in solar cell manufacturing. Many techniques have been employed to achieve this goal, e.g. plasma etching, grinding, wet etching, etc. Unfortunately, these methods have the disadvantages of being very expensive, single cell process and consume a lot of dangerous chemicals. This work introduces a simple technique for edge shunt isolation, by making a mechanical grove through the emitter and uses it as back contact. It was found that with this technique the solar cell open circuit voltage can be increased by an amount up to 200 mV. This technique is simple, cheap and environmentally friendly in the sense that it uses no chemicals**.

## I. Introduction

SOLAR cells always show degradation in their characteristic parameters (Voc, Isc, FF and efficiency) due to the defects introduced by multistep processing or material quality.

Shunts are one of the loss mechanisms which are caused by defects [1], they are divided into two categories volume and edge shunts. Volume shunts which account for 20% [2] of this loss mechanism are hard to remove without destroying the cell.

Edge shunts accounts for 80% of the loss mechanism and can be removed by many techniques. The first is plasma etching where the cells are coin stacked, then edges are etched using plasma, the second is wet chemical etching in which the edges are etched using KOH [3] and in the third method cell edges are removed by cutting leading to a reduction in the cell active area [4].

This work introduces a very simple cost effective technique for edge isolation, consists of making mechanical grove in the emitter at the cell front surface and uses it as a back contact.

## II. Theory

It has been found that the solar cell can be described by a two diode model [5] in a circuit consists of two parts, the cell body and edges as shown in fig.(1-a). The first part represents the main body which contains a series resistance Rs , shunt resistance Rsh, current source IL and a diode current I1 represents recombination current from sources other than the cell edges, given by

$$I_1 = I_{01} [\exp(qv/KT) - 1] \qquad (1)$$

where q is the electron charge, k is the Boltzman constant, and T is the absolute temperature.

The second part, which represents the edges, and consists of a diode current given by

$$I_E = I_{0E} [\exp(qv/m_E KT) - 1] \qquad (2)$$

where $I_E$ stands for edge recombination currents. The two parts are connected by a resistance $R_E$. $I_{0E}$ is the diode dark saturation current, and m is the diode ideality factor.

In this approach the edges are assumed to be eliminated, resulting in a low recombination currents, and hence the two diode model in fig. (1-a) can be converted to the well-known one diode model fig. (1-b).

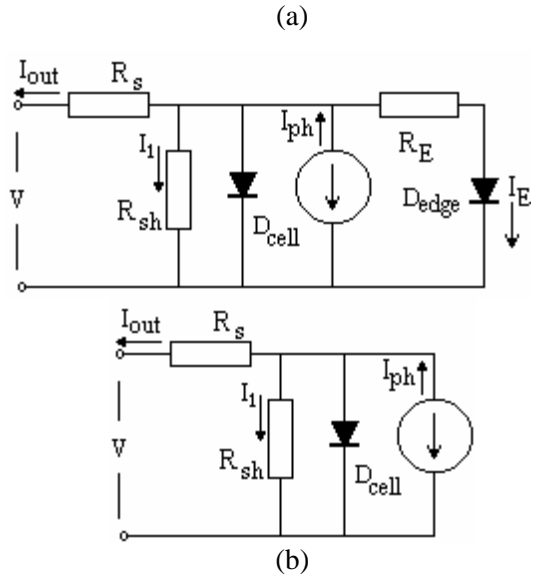

(a)

(b)

Fig 1 Equivalent circuit of the solar cell (a) before and (b) After, removing shunts [3].

### III. EXPERIMENT

The starting material is the commercially available p-type (111) CZ-Si wafers with an average resistivity around 1.06 cm. Prior to diffusion, the wafers are RCA cleaned [6]. Emitter diffusion has been carried out using Spin-on-Dopant (Filmtronics P509), which has been spunned and cured at 120 oC for 15 minutes [7].

Drive in diffusion is performed in quartz tube furnace at 1050 oC for 40 minutes. Finally the wafers are etched in buffered HF, then groove is made using a diamond cutter as in Fig.(2), finally Aluminum is deposited to form the front and back contacts.

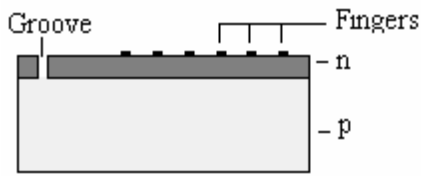

Fig. 2 Schematic representation of the solar cell.

### IV. EDGE SHUNT ISOLATION

As mentioned the technique is very simple, it involves making grove at the cell surface, i.e. bypassing the emitter leading to an isolation of the shunts at the edge. The main idea is to eliminate the shunts between the Emitter and the Base. Groves have been made using diamond cutter; the hardness of the cutter material guarantees there will be no contamination especially as there is no further high temperature processing. Voc measurements (Fig 2) through the grove show that Voc has improved by an amount up to 200 mV.

### V. RESULTS

Measurements of cell open circuit voltage have been made before and after grooving, results are shown in fig (3), and indicated by (Before and After), an increase in Voc by at least 100 mV can be seen which reflects the efficiency of the technique in improving Voc. The difference in the Voc increment is owed to the fact that, the weight of the edge shunts is not the same for all cells, a result which could be realized by noticing the differences in the open circuit voltage for each cell.

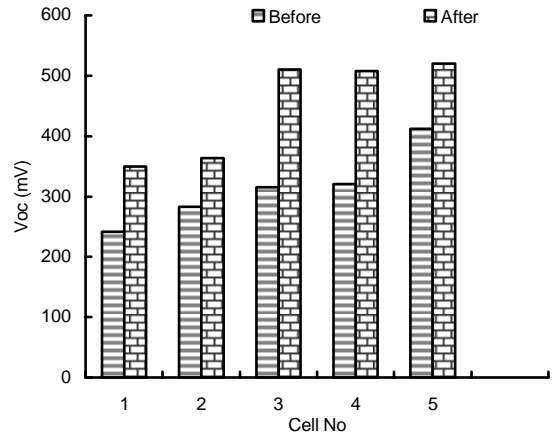

Fig. 3 Solar cell open circuit voltage before and after grooving.

### VI. CONCLUSION

A simple technique of edge shunt elimination has been introduced which can be easily applied to improve the solar cell open circuit voltage and consequently it is efficiency. It can be applied without losses in the cell active area leading to current losses or the need to use advanced techniques.

The method is found to be efficient not only for defected solar cell but for good solar cells as well, and can be implemented easily in mass production.

ACKNOWLEDGEMENTS

The research grant of the MOSTE is greatly appreciated. I. Hamammu would like to thank Garyounis University, Benghazi – Libya for their financial support. Thanks are also to Mr. Jamil Kassim.